\documentclass[aps,floats,twocolumn,epsf,prb,showpacs]{revtex4-1}

\usepackage{latexsym,amsmath,amssymb,bm,graphicx,amsfonts,bbm}
\usepackage{graphicx,color}
\usepackage{psfrag}
\usepackage{braket}
\usepackage{epsfig}
\makeatletter\def\Dated@name{}\makeatother

\newcommand{\UU}{{\cal U}}
\newcommand{\HH}{{\cal H}}

\newcommand{\VV}{{\cal V}}

\newcommand{\C}{\mathbb{C}}
\newcommand{\one}{\mathbbm{1}}
\newcommand{\zero}{0}
\DeclareMathOperator{\tr}{Tr}

 \renewcommand{\onlinecite}{\cite}

\begin{document}

\title{Dipole matrix element approach vs.\ Peierls approximation for optical conductivity}
\author{P. Wissgott$^{1}$, J. Kune\v{s}$^2$, A. Toschi$^1$ and K. Held$^1$}
\affiliation{$^1$ Institute for Solid State Physics, Vienna University of Technology,
1040 Vienna, Austria\\
$^2$ Institute of Physics, Academy of Sciences of the Czech republic, Cukrovarnick\'a 10,
Praha 6, 162 53, Czech Republic
}
\date{\today}

\begin{abstract}
We develop a computational approach for calculating the optical conductivity
in the augmented plane wave basis set of Wien2K and apply it for  thoroughly  comparing the full dipole matrix element calculation and the Peierls approximation.
The results for SrVO$_3$ and V$_2$O$_3$ show that the Peierls approximation,
which is commonly used in model calculations, works well for optical transitions between the $d$ orbitals. In a typical transition metal oxide, these transitions are solely responsible for the optical conductivity at low frequencies.
The Peierls approximation does not work, on the other hand, for optical transitions between $p$- and $d$-orbitals which usually  became important at  frequencies 
of 
a few eVs.
\end{abstract}

\maketitle
Much of our knowledge about solid state systems comes from their
response to small electro-magnetic perturbations. A broad range of
techniques has been developed to probe the nature of ground states
in elastic scattering experiments and the excitations in inelastic scattering
or absorption experiments. It is usually a combination of several
experimental techniques as well as theoretical calculations which
allow us to draw a complete picture of a given material.
Among those, the optical spectroscopy plays an important role \cite{opticsgeneralreview},
complementing the photoemission spectroscopy (PES) which is easier to
access and interpret in most theories. Probing the particle-hole
excitations averaged over the Brillouin zone, the optical spectra contain
a different and less detailed information about the system than
angle-resolved photoemission spectra. 
The main asset of the optical spectroscopy, however, 
is its robustness: Unlike PES, it does not suffer from
surface effects. Moreover, unlike transport measurements, the optical conductivity  is not critically affected by impurities or disorder: Optical transitions cannot simply disappear, but 
can
only be shifted to different energies, which is expressed by the sum-rule for optical conductivity \cite{Mahan1990}.

Calculations of optical spectra from first principles are well established within 
the effective non-interacting electron theories  \cite{Sole} for weakly correlated materials
such as the local density approximation (LDA) \cite{Jones89a} to the density functional theory.
The many-body perturbation theory on the GW level \cite{Hedin} and its two-particle extensions using the
Bethe-Salpeter equation \cite{Rohlfing1998} have been successful in describing the excitonic physics in semiconductors. 
The situation is different in the field of strongly correlated electron systems.
Although the optical measurement on these materials proved very useful for investigation
of metal-insulator transitions or mass renormalization effects,
material specific theoretical investigations are rather rare~\cite{Tomczak2009,Tomczak2009a,Tomczak_thesis,Oudovenko2004,Haule2005,Tomczak2012}. This is perhaps not surprising
given the fact that calculation of one-particle spectra is already a formidable
challenge. 

In the past decade the dynamical mean-field theory (DMFT) \cite{Metzner89a,Georges92a,Georges96a}
combined with first-principles bandstructures (LDA+DMFT) \cite{LDADMFTrev,Held2007} showed considerable
power to describe correlated materials. This theory 
allows for an accurate description of the local (intra-atomic) dynamics,
while the inter-atomic effects are treated on the static mean-field level.
Importantly, DMFT is not restricted to a particular energy scale and thus
allows for the simultaneous description of quasiparticles on meV scale and
atomic excitations on the eV scale, which is crucial to capture the spectral
weight transfers in the optical spectra.
First optical calculations with DMFT were performed for the single-band Hubbard model.
It was shown that the local approximation of DMFT leads to vanishing of the vertex corrections
to the optical conductivity \cite{Pruschke93a,Pruschke93b,Bluemer,note_vertexcorr}. This means that the electron and hole created in the 
process of optical excitation behave independently and thus the Green's function of
the electron-hole pair is a product of two one-particle Green's functions.
This is not necessarily true in multi-band models, except for the case of degenerate bands. However, the dipole selection rules at optical frequencies
typically forbid creation of an electron-hole pair on the same atom
and thus the vertex corrections may be neglected also in this case, an approximation
we also adopt throughout this work. Note that for inelastic x-ray scattering experiments 'optical' transitions
with finite momentum transfer allow formation of strongly bound local electron-hole pairs, excitons. The vertex
corrections in this case are substantial. A typical example are the crystal-field
$d$-$d$ excitations deep in the optical gap of transition metal oxides; see Ref.~\onlinecite{Hansmann11} for a V$_2$O$_3$ calculation.

While the formal framework for calculating the optical conductivity within
the above approximations is well established, the numerical implementation poses several 
challenges: i) $\mathbf{k}$-space integration, ii) determination of the optical transition
amplitudes and inclusion of states in a broad energy window, iii) evaluation
of optical spectra for real frequencies, which is an additional problem arising
for particular numerical techniques, such as quantum Monte-Carlo simulations, used to solve
the DMFT equations. Different strategies for dealing with these
issues are possible. In this article, we present an implementation based
on the Wannier functions formalism and a direct calculation of the transition amplitudes
from the one-particle wave functions. We compare our results to the so-called
Peierls approximation, which relies on the {\bf k}-derivatives of the effective
low-energy Hamiltonian of the systems considered and discuss their relationship. We analyze specifically two well-known correlated oxides, SrVO$_3$ and V$_2$O$_3$, as archetypes, and compare our results to the available experimental data.

The outline of the paper is as follows:
In Section
\ref{Sec:ComputationalDetails}, we give details on the LDA+DMFT
calculation. In  Section
\ref{Sec:dipole} the technical details of
the dipole matrix element caculation are discussed.
 Section
\ref{Sec:Peierls} discussed the relationship to
the Peierls approximation, which is popular for lattice models.
 Sections
\ref{Sec:SrVO3} and 
\ref{Sec:V2O3} present the results for  SrVO$_3$ and V$_2$O$_3$, respectively.
Finally, Section \ref{Sec:summary} summarizes the main findings.

\section{LDA+DMFT with Wannier orbitals}\label{Sec:ComputationalDetails}
The DMFT equations are naturally formulated in terms of fermionic creation and annihilation operators on a lattice,
a formulation which assumes an underlying set of localized orthogonal orbitals.
Our starting point are the LDA Bloch states $\psi_{n\mathbf{k}}$ and corresponding band-energies $\varepsilon_{nk}$ 
calculated with the full-potential linear augmented plane waves~(LAPW) program Wien2k.~\cite{Blaha1990,footnotekmesh} 
Depending on the specific material considered, we choose an energy window defined by the lower and upper band indices $n_\text{min}\le n \le n_\text{max}$ and transform
the states $\psi_{nk}$ from this window to real-space Wannier orbitals~\cite{Wannier1937} localized around lattice sites ${\bf R}$:
\begin{align}\label{Eq:Fouriertransform}
   \ket{w_{m\mathbf{R}}} = \frac{1}{N_\mathbf{k}}\sum_{n\mathbf{k}} e^{i\mathbf{k}\mathbf{R}}U_{mn}(\mathbf{k})\ket{\psi_{n\mathbf{k}}}
\end{align}
where $U(\mathbf{k})$ are unitary matrices defined throughout the Brillouin zone and $N_\mathbf{k}$ is the number of $\mathbf{k}$-points. Using wien2wannier~\cite{Kunes2010} and wannier90~\cite{Mostofi2008} the matrices $U(\mathbf{k})$ that lead 
to maximally localized Wannier functions are found. Construction of the single-particle part of the effective Hamiltonian
is completed by rotation of the LDA Hamiltonian into the Wannier basis
\begin{align}\label{Eq:H}
   H^{W}_{mm'}(\mathbf{k}) = \sum_n U_{mn}^{+}(\mathbf{k})\varepsilon_{n\mathbf{k}}U_{mn'}(\mathbf{k}).
\end{align}
Finally the on-site interaction is added to the Hamiltonian. Another important input data required for the DMFT calculation, are the local Coulomb repulsion parameters which define the 
term
to be added to $H^{W}_{mm'}(\mathbf{k})$ in the Wannier basis: the intra-orbital local repulsion $U$, inter-orbit local interaction $V$ and the exchange parameter $J$. 
In principle, this input should be computed from the underlying LDA data, with constrained LDA~\cite{Held2007} or constrained random phase approximation~\cite{Aryasetiawan04}. 
However, since identifying $U$, $V$, and $J$ for SrVO$_3$ or V$_2$O$_3$ 
is not the aim
of this work, we adapted these values from the literature~\cite{Nekrasov,Held2001,Keller2004,Poteryaev2007}. In the case  V$_2$O$_3$ we have chosen a slightly lower value of $U$ than in Refs.\onlinecite{Held2001,Keller2004,Poteryaev2007}, which ensures the best agreement with XAS\cite{Rodolakis2010} and optical experiments\cite{Lupi2010}, according to considerations reported in Ref. \onlinecite{Toschi2010}.  It is important to notice though that the Coulomb parameter as well as the DMFT itself in general depend on the chosen basis set of Wannier function, which especially becomes important if the choice of Wannier orbitals is not as straightforward as in the case of SrVO$_3$ or V$_2$O$_3$ below.
In both cases the actual DMFT calculation has been done in the $t_{2g}$ 
subspace. For the optical conductivity, the DMFT   $t_{2g}$ Green function
was then supplemented with the LDA Green function for the other orbitals.

 Once the effective LDA Hamiltonian is set up, the DMFT equations are solved numerically using  quantum Monte Carlo simulations \cite{HirschFye} with an imaginary time discretization of $\Delta \tau=0.1$ eV$^{-1}$ for SrVO$_3$ and $\Delta \tau=0.125$ eV$^{-1}$ for V$_2$O$_3$, respectively, to obtain the one-particle self-energy which serves as the many-body input for evaluation of the optical conductivity.
Similar LDA+DMFT approaches based on augemted plane waves
can be found in Ref.~\onlinecite{Lechermann2006,Amadon2008,Aichhorn2009}.\\
From the LDA+DMFT self energy $\Sigma(i\omega_m)$ for a temperature $T$ 
at
the Matsubara frequencies $\omega_m=\pi(2m+1)T$, we obtain $\Sigma$ on real frequencies via the procedure described in Ref.~\onlinecite{Nekrasov2006}: Starting from the imaginary-time Green's function $G(\tau)$ measured in Monte-Carlo, the $\mathbf{k}$-integrated spectrum $A(\omega)$ is calculated by the maximum entropy method~(MEM, see Ref.~\cite{MaxEnt}). Afterwards, the local Green's function $G_{MEM}(\omega)$ for real frequencies is found by applying Kramers-Kronig relations. Finally, we fit $\Sigma(\omega)$ such that the Green's function obtained by direct $\mathbf{k}$-summation, i.e. $G(\omega)= 1/N_\mathbf{k} \, \sum_\mathbf{k} [\omega-H_W(\mathbf{k})-\Sigma(\omega)]^{-1}$, matches the one from the maximum entropy method.

\section{Linear response for the optical conductivity}\label{Sec:Method}
\subsection{Dipole matrix element approach}\label{Sec:Dipolematrixapproach}\label{Sec:dipole}
The regular part of the  optical conductivity is obtained via the standard Kubo's formula in linear response~\cite{Mahan1990}
\begin{align}\label{Eq:KuboFormula}
 \sigma_{\alpha\beta}(\omega) =  \lim_{\mathbf{q}\rightarrow 0} \mbox{Re}\left(\frac{1}{\omega V}\int\, dt\ e^{i\omega t}\braket{\left[j_{\alpha}(\mathbf{q},t),j_{\beta}(-\mathbf{q},0)\right]}\right)
\end{align}
where $V$ is the unit cell volume, $j_\alpha(\mathbf{q},t)$ is the $\mathbf{q}$-momentum paramagnetic current in $\alpha$ direction, $\hbar=1$ and $q\rightarrow0$ is the dipole approximation. 
Expressing Eq. \eqref{Eq:KuboFormula} via the Lindhardt bubble and Matsubara formalism~ (thus omitting vertex corrections, consistently with the discussion above), we include internal degrees of freedom describing optical excitations~\cite{Mahan1990}: initial~(final) frequency $\omega$,($\omega+\Omega$), reciprocal vector $\mathbf{k}$ and $N= n_{max}-n_{min}$ orbital degrees of freedom participating in optical transitions. Altogether, we obtain 
the following expression for the real part of the optical conductivity
\begin{equation}\label{Eq:Optical conductivity}
\begin{split}
 \sigma_{\alpha\beta}(\omega) =& \frac{2\pi}{V}\sum_{\mathbf{k}}\int {\rm d} \omega'\frac{f(\omega')-f(\omega'+\omega)}{\omega}\\
&\times\tr\left[A(\mathbf{k},\omega')v_W^{\alpha}(\mathbf{k})A(\mathbf{k},\omega'+\omega)v_W^{\beta}(\mathbf{k})\right]
\end{split}
\end{equation}
where $f$ is the Fermi function, $v_W^\alpha(\mathbf{k})=U(\mathbf{k})v^{\alpha}(\mathbf{k})U^+(\mathbf{k})$ are rotated matrix elements of the momentum operator $v^{\alpha}_{nm}(\mathbf{k})=-i \braket{\psi_{n\mathbf{k}}|\nabla_\alpha|\psi_{m\mathbf{k}}}/m_e$, $1\le,n,m\le N$, the elementary charge $e=1$, and $A(\mathbf{k},\omega)=-{\rm Im} G(\mathbf{k},\omega)/\pi$ is the generalized spectral function with the Greens function 
\begin{align}\label{Eq:Greensfunction}
 G(\mathbf{k},\omega) =  \left[(\omega+\mu)\one-H_{W}(\mathbf{k})-\Sigma(\omega)\right]^{-1}.
\end{align}
Here, $\mu$ denotes the chemical potential, $H_W(\mathbf{k})\in\C^{N\times N}$ the~(non-interacting) Hamiltonian in the Wannier orbital basis and $\Sigma(\omega)\in\C^{N\times N}$ the self energy from the LDA+DMFT calculation.\\
For an efficient and accurate $\mathbf{k}$-quadrature of Eq.~\eqref{Eq:Optical conductivity}, we use a tetrahedral-mesh integration. To resolve regions in $\mathbf{k}$-space with larger integration error we adaptively refine the tetrahedra in these domains. Furthermore, the symmetry operations of the unit cell are applied such that the integrand of Eq~\eqref{Eq:Optical conductivity} has to be evaluated only at $\mathbf{k}$-points within the reduced wedge of the Brillouin zone.\\
The computation of the momentum matrix elements 
\begin{align}\label{Eq:DipoleMatrix}
 v_{nm}(\mathbf{k})=-i \frac{\braket{\psi_{n\mathbf{k}}|\nabla|\psi_{m\mathbf{k}}}}{m_e}
\end{align}
which 
are
in the following also denoted as dipole matrix requires their evaluation in term of the underlying LAPW basis set~\cite{Ambrosch-Draxl2006}. It is thus~(to our knowledge for the first time) possible to combine a full potential LAPW dipole matrix with Wannier-functions-based DMFT algorithms for the computation of transport and optical properties~(for different approaches see e.g. Refs.~\onlinecite{Ambrosch-Draxl2002,Peschel2005}). Note that the surveyed workflow is not limited to the use of a DMFT self energy $\Sigma(\omega)$, but can be easily generalized for other, even $\mathbf{k}$-dependent self energies $\Sigma(\mathbf{k},\omega)$. In such cases, however, the inclusion of vertex corrections to the bubble term becomes usually necessary \cite{Gunnarsson2010}. \\
In addition to transitions within 
Hilbert space
of the low energy model, the 
present
approach also allows 
inclusion of 
higher energy bands.
This can be achieved by 
enlarging
the transformation matrices $U(\mathbf{k})$ and, consequently, the Hamiltonian $H_W(\mathbf{k})$
\begin{align}\label{Eq:LargMatrix}
  \UU(\mathbf{k})=&\left[\begin{array}{ccc}\one &\zero& \zero \\
                                \zero &U(\mathbf{k})& \zero \\
                                 \zero &\zero &\one\end{array} \right],\\ 
\HH(\mathbf{k})=&\left[\begin{array}{ccc}E^{(1)}(\mathbf{k}) &\zero& \zero \\
                                \zero &H_W(\mathbf{k})& \zero \\
                                 \zero &\zero &E^{(2)}(\mathbf{k})\end{array} \right]
\end{align}
with diagonal $E^{(1)}_n(\mathbf{k}), E^{(2)}_n(\mathbf{k})=\varepsilon_{n\mathbf{k}}$ for $n<n_{min}$~($n>n_{max}$). Note that though the $\UU$ and $\HH_W$ are block-diagonal, the corresponding dipole matrix $\VV(\mathbf{k})=\UU(\mathbf{k}) v^\alpha(\mathbf{k}) \UU^+(\mathbf{k})$
is 
not. 
Inserting $\UU,\HH,\VV$ into Eq.\eqref{Eq:Optical conductivity} and \eqref{Eq:Greensfunction}, we thus also take transitions between the Wannier orbitals and Bloch states outside of the low energy model into account.
%
\subsection{Peierls approximation}\label{Sec:Peierlsapproximation}\label{Sec:Peierls}
For many-body calculations of lattice models, it is common practice to
determine the optical conductivity by the Peierls approximation(PA) \cite{Peierls1933}.  The PA approximates the group velocities directly from 
 the hopping elements and, for non-Bravais lattices, from the atomic positions in the unit cell \cite{Tomczak_thesis,Tomczak2009,Tomczak2009a}. If one wants to go beyond the PA, however, one needs to know the
underlying continuum description for calculating the dipole matrix elements.
The idea of the PA is a gauge transformation of the electro-magnetic potential $\mathbf A$
which disregards the inner orbital structure (an orbital will get a different gauge factor at different positions) and assumes a single gauge factor which only depends on the lattice site. This 
is reflected 
in a modified hopping amplitude
$t_{{\bf R} m; {\bf R'}n} \rightarrow t_{Rm;R'n}\exp[i{\bf A} ({\bf R}-{\bf R'})/c ]$  \cite{Bluemer,Tomczak_thesis} between sites  ${\bf R}$ and  ${\bf R'}$.

In the following we discuss the corrections to the PA, emerging from the exact continuum description in the  Wannier orbitals basis, cf. Ref.~ \onlinecite{Tomczak_thesis}.
Using the operator identity $-\tfrac{1}{m}\nabla=[H_0,\mathbf{r}]$, where $H_0$ is the one-particle
part of the Hamiltonian we can write the momentum matrix element as
\begin{equation}
\begin{split}
 &\frac{-1}{m_e}\braket{w_{m\mathbf{k}}|i\nabla|w_{m'\mathbf{k}}} = \frac{i}{N}\sum_{\mathbf{R},\mathbf{R}'}\! e^{i\mathbf{k}(\mathbf{R}'-\mathbf{R})}
\braket{w_{m\mathbf{R}}|[H_0,\mathbf{r}]|w_{m'\mathbf{R}'}}\\
  &=\frac{i}{N}\sum_{\mathbf{R},\mathbf{R}'} e^{i\mathbf{k}(\mathbf{R}'-\mathbf{R})} \biggl[(\mathbf{R}'-\mathbf{R})
\braket{w_{m\mathbf{R}}|H_0|w_{m'\mathbf{R}'}}\\
  &+ \braket{w_{m\mathbf{R}}|H_0(\mathbf{r}-\mathbf{R}')|w_{m'\mathbf{R}'}}-\braket{w_{m\mathbf{R}}|(\mathbf{r}-\mathbf{R}) H_0|w_{m'\mathbf{R}'}}\biggr]. \label{Eq:PAcorr1}
\end{split}
\end{equation}
The first term equals $\nabla_{\mathbf{k}} H(k)$, the PA,
and can be obtained without explicit 
knowledge of the orbital, e.g. from an empirical tight-binding Hamiltonian. 
The remaining two terms can be further analyzed 
by noting that the Wannier functions form a complete 
eigenbasis of $H_0$. Hence we have 

\begin{equation}
 -\frac{1}{m_e}\braket{w_{m\mathbf{k}}|i\nabla|w_{m'\mathbf{k}}} = \nabla_\mathbf{k} H_{mm'}(\mathbf{k})+ \mathcal{C}_{mm'}(\mathbf{k}),
\end{equation}
where
\begin{equation}
\begin{split}
  &\mathcal{C}_{mm'}(\mathbf{k})= \frac{i}{N}\sum_{\mathbf{R},\mathbf{R}'} e^{i\mathbf{k}(\mathbf{R}'-\mathbf{R})} \times \\ \sum_{\mathbf{L},p} 
  & \biggl[\braket{w_{m\mathbf{R}}|H_0|w_{p\mathbf{L}}}\braket{w_{p\mathbf{L}}|\mathbf{r}-\mathbf{R}'|w_{m'\mathbf{R}'}}\\
  &- \braket{w_{m\mathbf{R}}|\mathbf{r}-\mathbf{R}|w_{p\mathbf{L}}}\braket{w_{p\mathbf{L}}|H_0|w_{m'\mathbf{R}'}}\biggr].
\label{Eq:PAcorr}
\end{split}
\end{equation}
Let us first discuss the corrections for a single atom in the unit cell. These can be classified as follows\\
 (i) {\em Intra-atomic dipole transitions}:
Terms in Eqs . (\ref{Eq:PAcorr1})  and (\ref{Eq:PAcorr}) with $\mathbf{R}=\mathbf{R}'$ yield together $\braket{w_{m\mathbf{R}}| [H_0, \mathbf{r}]|w_{m'\mathbf{R}}}$, i.e., the atomic-dipole elements with the only difference being that $H_0$ is the one-particle Hamiltonian of the solid and not of the atom. These local transitions generally require different angular momenta for $m$ and $m'$ orbitals and are hence at a higher energy. They cannot be described by the PA which only considers a single gauge factor for the atom or site.\\
(ii)  {\em Dipole transition mediated hopping}: 
For  $\mathbf{R}'=\mathbf{L}\neq\mathbf{R}$, the first term of
 Eq.\ (\ref{Eq:PAcorr}) consists of  a hopping integral $\braket{w_{m\mathbf{R}}|H_0|w_{p\mathbf{R}'}}$ and a local dipole transition $\braket{w_{p\mathbf{R}'}|\mathbf{r}-\mathbf{R}'|w_{m'\mathbf{R}'}}$. This is similar as  intra-atomic dipole transitions, however now we obtain a $\mathbf{k}$-dependence which was absent for (i).
Note, the same is obtained for  the second term of  Eq.\ (\ref{Eq:PAcorr}) in the case
 $\mathbf{R}=\mathbf{L}\neq\mathbf{R}'$.\\
(iii)  {\em Inter-atomic dipole transitions}:
For  $\mathbf{R}=\mathbf{L}\neq\mathbf{R}'$, the first term of
 Eq.(\ref{Eq:PAcorr}) consists of a local Wannier matrix element
$\braket{w_{m\mathbf{R}}|\hat{H}_0|w_{p\mathbf{R}}}$ and an
inter-atomic dipole transition
 $\braket{w_{p\mathbf{R}}|\mathbf{r}-\mathbf{R}'|w_{m'\mathbf{R}'}}$.
A similar term with a minus sign is obtained for  the second term of  Eq.(\ref{Eq:PAcorr}) in the case
 $\mathbf{R'}=\mathbf{L}\neq\mathbf{R}$. If the orbitals are locally orthogonal, only the local on-site energies survive, and we only get a contribution if there is  a crystal field splitting of the orbitals. 
\\
(iv)   {\em Further corrections} arise if all lattice positions
 $\mathbf{R}$, $\mathbf{L}$ and  $\mathbf{R}'$ are different in Eq.(\ref{Eq:PAcorr}). 
In this case we have a combination of an inter-atomic dipole element and
a hopping term.\\
If the orbitals are more localized, i.e., exponentially decaying
between the atoms, both the hopping element and the   inter-atomic dipole element are affected by  this exponential suppression.  
Hence the terms (iv), which contain {\sl two} such exponentials, are more strongly suppressed than  the hopping amplitude itself (which enters (ii) and (iii) as well as the PA) since it only contains {\sl one} exponential factor.
The leading term in the ``localized'' limit is (i), which only involves local transitions.

From these general considerations, the PA appears a rather unjustified approximation. In fact, even in the limit of more localized orbitals only the terms (iv) get suppressed. However, in specific cases of interest the PA may be justified. For instance, terms (i)-(iii) become only relevant if the orbitals are 
(a) affected by 
a large crystal
field splitting or (b) 
of
a different angular momentum, which typically also means 
large excitation
energies.  Hence for transitions below 
this
energy, 
e.g. the Drude peak,
PA is expected to work, at least for sufficiently localized orbitals.

The situation becomes a 
bit more involved in the case of
several atoms in the unit cell. Tomczak and Biermann \cite{Tomczak2009,Tomczak2009a} 
showed that the PA has to be generalized to include the hopping terms between
the atoms in the same unit cell, which are absent in 
 $\nabla_\mathbf{k} H_{mm'}(\mathbf{k})$.  However, also in this case, the same
correction terms (i)-(iv) as discussed above remain.

\section{Results}\label{Sec:Results}
\subsection{SrVO$_3$}\label{Sec:SrVO3}
Due to its simple cubic (perovskite) lattice structure and $3d^1$ 
electronic structure, SrVO$_3$   has been employed as a testbed for
 {\em ab initio} calculations such as LDA+DMFT. There is, on average,
 a single $d$ electron residing in three degenerate $t_{2g}$ bands that
 cross the Fermi energy $E_F$. These $t_{2g}$ bands  are well separated
 by a gap from the oxygen $p$ bands below and the $e_g^{\sigma}$
 orbitals above. This situation makes the electronic structure of
 SrVO$_3$ particularly simple.

The photoemission spectra~\cite{Eguchi2006,Inoue1994,Yoshida2010,Sekiyama2004} show a well 
developed lower Hubbard 
below $E_F$ band and a pronounced quasiparticle peak around $E_F$;
an upper Hubbard band is found, on the other side, in x-ray absorption
experiments. \cite{Inoue1994}
The quasiparticle peak  is renormalized (narrowed) by  a factor
 of about $2$ compared to the overall LDA $t_{2g}$  bandwidth.
 \cite{Sekiyama2004} This is in good agreement with 
LDA+DMFT calculations \cite{Nekrasov,Sekiyama2004} in which the
 interaction parameters have been determined from constrained 
LDA calculations \cite{cLDA}. Essentially the same one-particle 
spectrum has also been obtained in subsequent LDA+DMFT calculations~(among others see Refs.~ \onlinecite{Pavarini2004,Liebsch2003,Nekrasov2005,Lechermann2006,Zhang2007,Amadon2008,Werner2010,Trimarchi}); 
 and various Wannier function projection schemes
have been tested for this prototypical material~(among others see Refs.~\onlinecite{Anisimov2005,Solovyev2006,Pavarini2005,Miyake2008,Freimuth2008}).
 SrVO$_3$ is also the materials where kinks in 
strongly correlated electron systems,
 abstain from any external bosonic degrees of freedom and
anti-ferromagnetic spin-fluctuations have been discovered.
 \cite{Nekrasov2005,Byczuk2007} Similar structures can also be identified  in angular resolved
 photoemission spectra. \cite{Yoshida05}
As the optical conductivity averages (integrates) however over 
different $\mathbf{k}$-points, such fine structures are hardly discernible 
in this physical quantity.\cite{Toschi12} Experimentally, the 
optical conductivity shows 
a Drude peak  and additional features above 2$\,$eV when transitions 
between Hubbard and quasiparticle peak become relevant~(among others see Refs.~\onlinecite{Makino1998,Mossanek2009}).

\subsubsection{Spectral properties}\label{Sec:ResultsSrVO3spectra}

\begin{figure}[t]
  \centering
 \includegraphics[width=7cm]{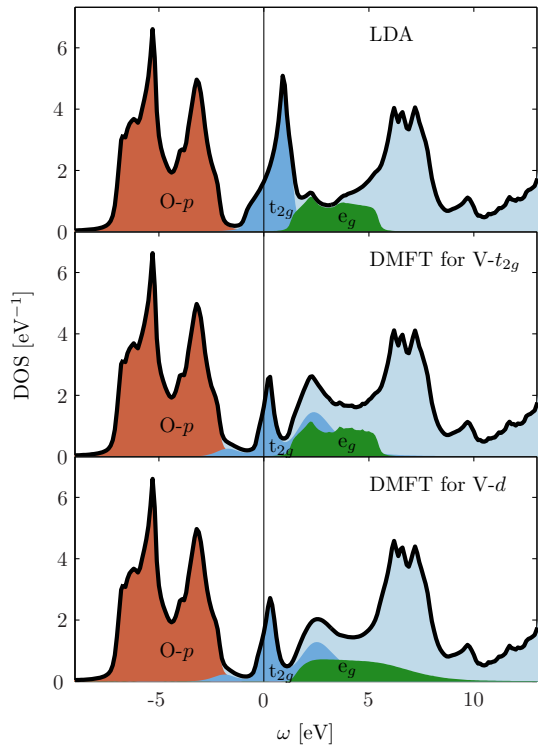}
\caption{(Color online) Non-interacting partial density of states of SrVO$_3$~(results abbreviated as LDA) compared to DMFT spectra at $T=1160 K$. Two DMFT basis sets were employed: first, the three orbital t$_{2g}$ basis with parameters $(U,J,V)=(5.05,0.75,3.55)$ eV; second, the entire V d manifold with 
the same interaction parameters for all orbitals.
    \label{Fig:SrVO3_dos1}}
\end{figure}

The LDA density of states~(DOS) for SrVO$_3$ used in our analysis can be found in Fig.~\ref{Fig:SrVO3_dos1}~(top panel), where the three partial DOS contributions V-$t_{2g}$, V-$e_{g}$ and O-$p$ are highlighted~(we use $a=3.84$~\AA~as unit cell lattice parameter for the cubic perovskite). From the LDA data, we obtained three different Wannier projections: First, just the V-$t_{2g}$ manifold was mapped onto three Wannier orbitals~(in the following abbreviated as $P1$). Second, we also included the two additional bands with predominant V-$e_{g}$ character and thus describe the full V-$d$ manifold~($P2$). Finally, we also take into account the O-$p$ bands which leads to a basis consisting of $14$ Wannier functions~($P3$).\\
In Fig.~\ref{Fig:SrVO3_dos1}, middle and lower panel, we plot the LDA+DMFT spectra computed with the Wannier basis sets $P1$ and $P2$, respectively. The parameters where adapted from Ref.~\onlinecite{Nekrasov}: local intra-orbital Coulomb repulsion $U=5.05$ eV, local inter-orbital repulsion $V=3.55$ eV, and local exchange $J=0.75$ eV. Compared to the LDA DOS, the $t_{2g}$ partial density of states is renormalized and the formation of lower and upper Hubbard bands can be observed as correlation effects are taken into account within the DMFT framework. In the case of $P2$, where all $5$ V Wannier orbitals are included in the DMFT, the $e_g$ orbitals remain completely unoccupied as in LDA leading to negligible correlation effects in these two orbitals~(see lower panel of Fig.~\ref{Fig:SrVO3_dos1}). We thus restricted the LDA+DMFT analysis for lower temperatures to $P1$, where only the $t_{2g}$ orbitals are described within DMFT.

\subsubsection{Comparison of the dipole matrix elements approach and the Peierls
 approximation}\label{Sec:ResultsSrVO3comparison}
\begin{figure}[t]
  \centering
  \includegraphics[width=7cm]{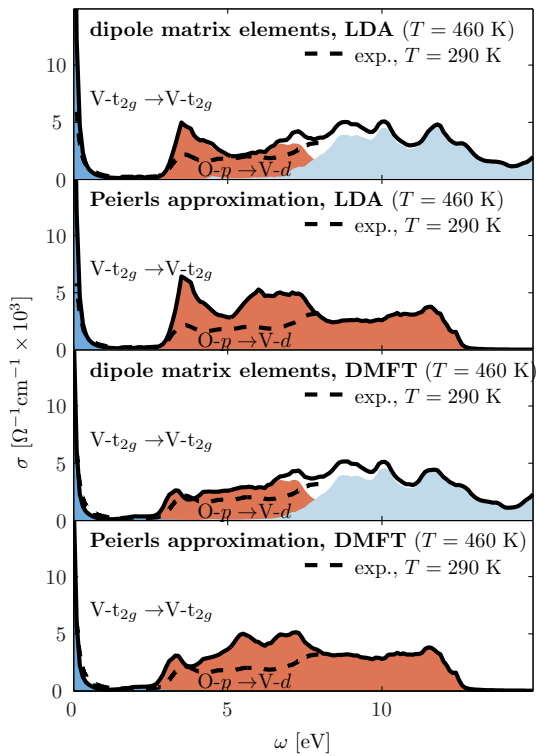}
\caption{(Color online) Optical conductivity of SrVO$_3$ calculated with dipole matrix elements and the Peierls approximation, respectively, compared to experiment~\cite{Makino1998}.}
    \label{Fig:SrVO3_opt1}
\end{figure}

In Fig.~\ref{Fig:SrVO3_opt1}, our main results for the optical conductivity of SrVO$_3$ are summarized. We compare four different calculations for the~(isotropic) optical conductivity $\sigma$ computed via Eq.\eqref{Eq:Optical conductivity} with the experimental data from Ref.~\onlinecite{Makino1998}: The uppermost panel shows $\sigma$ computed by use of the LDA Greens function~\eqref{Eq:Greensfunction}, where we fixed the broadening by setting $\Sigma=-0.04i$~[eV]  in Eq.(\ref{Eq:Greensfunction}), and employed the dipole matrix~\eqref{Eq:DipoleMatrix} as group velocities. The second panel of Fig.~\ref{Fig:SrVO3_opt1}, visualizes the optical conductivity $\sigma$ computed with the same Green function $G(\mathbf{k},\omega)$, but with the Peierls approximation~$\nabla_{{\bf k}}H_W(\mathbf{k})$ for the group velocities~(we are neglecting for this calculation the intra-unit-cell contributions as introduced by Tomczak~\cite{Tomczak2009a}). For the lower two panels, we inserted the DMFT self energy into the formula for the Greens function~\eqref{Eq:Greensfunction}. In particular, the third and the fourth panel of Fig.~\ref{Fig:SrVO3_opt1} show the LDA+DMFT results for $\sigma$ calculated with the dipole matrix and the Peierls approximation as group velocities, respectively. Note that the effect of taking a different temperature in the experiment~($T=290$ K) and in the calculations~($T=                                                                                                                                                                                                                                                                                                                                                                                                                                                                                                                                                                                                                                                                                                                                                                                                                                                                                                                                                                                                                                                                                                                                                                                                                                                                                                                                                           460$ K) is expected to be limited since in this temperature range SrVO$_3$ does not show a notable change in the electronic structure. The main consequence of lowering the temperature $T=460\rightarrow 290$ is the decreased electron-electron scattering within the coherent part of the electron spectrum which eventually leads to the Drude peak to become more pronounced while the inter-band contributions remain essentially unchanged.\\
In our analysis of the results, let us start investigating the qualitative effect of correlation on the optical spectra, i.e. comparing the upper two panels of Fig.~\ref{Fig:SrVO3_opt1} with the two lower ones. The renormalization of the $t_{2g}$ manifold surveyed in Fig.~\ref{Fig:SrVO3_dos1} leads to a smaller Drude weight in the LDA+DMFT panels of Fig.~\ref{Fig:SrVO3_opt1} and to a suppression of the prominent peak of $\sigma$ around $3.5$ eV predominately stemming from transitions from the occupied O-$p$ manifold to the unoccupied section of the V-$t_{2g}$ orbitals. The suppression of these two features is also seen in experiment. Additionally, the DMFT optical spectra in the lower two panels of Fig.~\ref{Fig:SrVO3_opt1} show the formation of a small satellite at $\sim2$ eV originating from transitions from the lower Hubbard band of the $t_{2g}$ orbitals to the unoccupied part of their coherent spectral peaks.\\
Comparing the dipole matrix approach with the Peierls approximation, i.e. the first with the second and the third with the fourth panel in 
Fig.~\ref{Fig:SrVO3_opt1}, indicates both reproduce low energy transitions in a similar way. Since the Drude peak in this material stems from intra-band excitations of the $t_{2g}$ bands, this implies that the Peierls approximation is sufficient to describe optical transitions in SrVO$_3$ as long as only well localized orbitals are participating. The case is different for the O-$p\rightarrow$V-$t_{2g}$ transitions, where a deviation in the range between $4$ eV and $13$ eV is clearly visible. Here, the dipole matrix matrix approach appears to be superior and $\sigma$ is much closer to the experimental results than the Peierls approximation, especially for the LDA+DMFT optical spectra~(see third and fourth panel of Fig.~\ref{Fig:SrVO3_opt1}). The reason for this behavior can be understood taking into account the more non-local nature of the O-$p$ orbitals and the deficiency of the Peierls approximation to describe optical transitions therein quantitatively correct.\\
In addition to deviations compared to the choice of group velocities, both LDA+DMFT results for the optical spectra deviate with experiment around $3.5$ eV at the onset of the O-$p\rightarrow$V-$t_{2g}$ transitions. Since LDA seems to describe this onset more accurately, we think that the reason for this behavior deduces from the fact that, including only the $t_{2g}$ orbitals in LDA+DMFT, we did not consider a double counting correction shifting the $t_{2g}$ orbitals relative to  the O-$p$ orbitals. A more complete approach would consider the O-$p$ within LDA+DMFT on the level of the Hartree approximation taking into account the double counting corrections more accurately. Then, the change of the $t_{2g}$ orbitals within LDA+DMFT would eventually shift the $p$ states to lower energies correcting the energy distance between the onset of the O-$p$ manifold to the~(now renormalized) peak in the $t_{2g}$ orbitals back to the LDA level.

\subsubsection{Sumrule analysis}\label{Sec:Sumrule}
\begin{figure}[t]
  \centering
 \includegraphics[width=7cm]{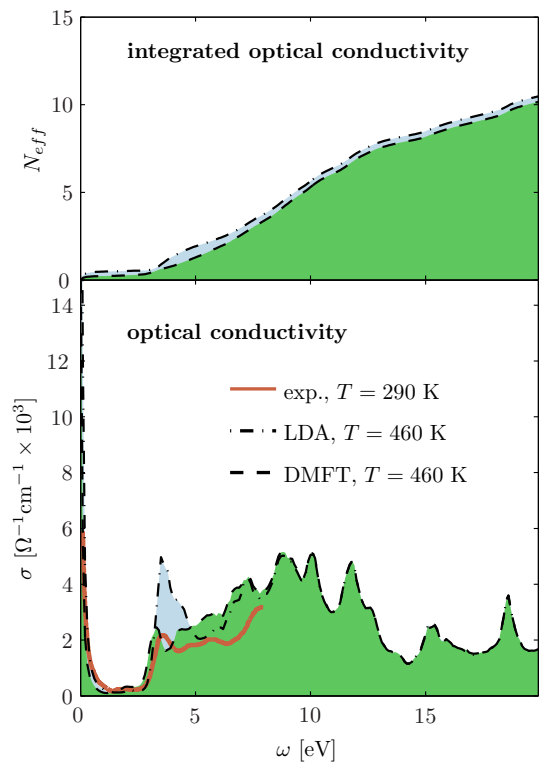}
\caption{(Color online) Optical conductivity of SrVO$_3$ comparing LDA, DMFT and experiment. The LDA and DMFT results were computed by the dipole matrix element approach~(bottom). The top panel shows the sumrule $N_{eff}(\Omega_c)$ from Eq.~\eqref{Eq:fsumrule} for $\sigma$ of the lower panel~(the experimental data are from Ref.~\onlinecite{Makino1998}.}
    \label{Fig:SrVO3_opt2}
\end{figure}
An important aspect associated to the  theoretical and experimental study of the optical spectroscopy is the analysis of the associated f-sum rule\cite{Mahan1990}. This is a direct consequence of charge conservation, stating that the integral over all frequencies of the optical conductivity is always proportional to the total electronic density $n_{tot}=N_{tot}/V$ of the system
 \begin{align}\label{Eq:fsumrule}
 \lim_{\Omega_c \rightarrow \infty} \frac{m_e}{\pi e^2} \int_0^\infty \, d\omega \,\, \sigma(\omega) =  \lim_{\Omega_c \rightarrow \infty} \frac{N_{eff}(\Omega_c)}{V} = \frac{N_{tot}}{V}.
  \end{align}
The importance of the f-sum rule, however, goes well beyond the verification of the charge conservation in LDA or LDA+DMFT calculations of optical spectra. The validation of Eq.~\eqref{Eq:fsumrule} in theoretical calculations as well as in experiment represents a rather academic but delicate issue, as it involves very different energy scales~(corresponding to optical transitions involving valence and core states). For further details about this issue, we refer the reader to Refs.\cite{Shiles1980,Smith00}.\\
More specific information can be extracted by the analysis of so-called partial or restricted optical sum-rules. They correspond to consider just a portion of the frequency integral in Eq.~\ref{Eq:fsumrule}, a typical case being a finite upper cut-off $\Omega_c$, and how this partial integral changes as a function of external parameters (e.g., temperature, magnetic field, etc...). This provides usually very important information about the energy balance associated, e.g,. with a phase transitions, as it has emerged from many experimental\cite{sumrule:exp} and theoretical analysis\cite{sumrule:theo} of integrated optical spectroscopic data of high-temperature superconducting cuprates, and most recently, by analyzing\cite{ciro2011} the non-Slater nature of the antiferromagnetic phase in the optical spectra of LaSrMnO$_4$\cite{goessling2008}.   

An example for the application of Eq.\ (\ref{Eq:fsumrule}) is reported in the upper panel of Fig. \ref{Fig:SrVO3_opt2}, where the growth of $N_{eff}$ with increasing frequency up to $\Omega_C=20$ eV is shown for the case of SrVO$_3$~(there are $19$ valence electrons included in our calculation). While a detailed analysis of the restricted sum rules for SrVO$_3$ goes beyond the scope of this work (for the analysis of the restricted sum rule in V$_2$O$_3$, see, e.g., Ref.\onlinecite{Baldassarre2008}), when comparing the integrated LDA and LDA+DMFT optical spectra of  Fig.~\ref{Fig:SrVO3_opt2}, we can note, for the latter case, a slight decrease of the values of $N_{eff}$ in the low-frequency region, which reflects, evidently, a correspondent reduction of the electronic mobility due to electronic correlations. At higher frequency, however, the LDA electron density value has been recovered within an accuracy of about $3$ \%.

\subsection{V$_2$O$_3$}\label{Sec:ResultsV2O3}\label{Sec:V2O3}
\begin{figure}[t]
  \centering
 \includegraphics[width=7cm]{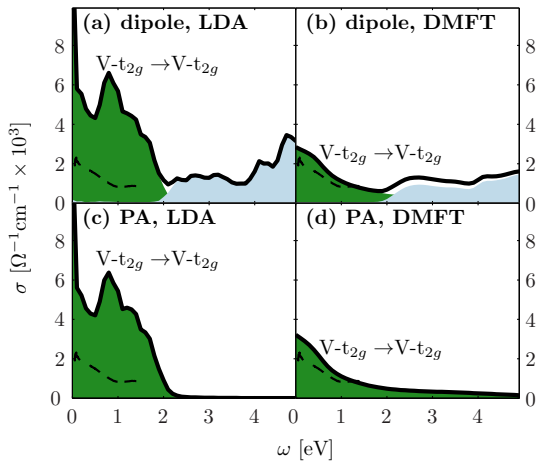}
\caption{(Color online) Optical conductivity of V$_2$O$_3$~($\alpha$-phase, metallic) at $T=460$ K calculated with dipole matrix elements~[a: LDA, b: DMFT for V-t$_2g$ with (U,J,V)=(4,0.7,2.6) eV] and the Peierls approximation~[c: LDA, d: DMFT for V-t$_2g$ with (U,J,V)=(4,0.7,2.6) eV], respectively, compared to experiment~(dashed line taken from Ref.\onlinecite{Rozenberg1995}).}
    \label{Fig:Parameter1}
\end{figure}


Vanadium sesquioxide V$_2$O$_3$ has been  
subject
of a considerable interest in condensed matter physics since the early Seventies (see e.g. Ref. \onlinecite{MacWhan1973}), as it represents one of the most evident realization of the Mott-Hubbard metal-to-insulator transition (MIT).
In fact, V$_2$O$_3$ can be relatively easily doped with Cr or Ti, and its phase diagram displays a clear first order transition between a paramagnetic metallic (PM) state (at low concentration of  Cr, or for Ti doping) and  a paramagnetic insulating (PI) at a higher level of Cr doping. Such a first order MIT, which emerges from a (simultaneous) lower temperature structural and magnetic transition and ends up at higher $T$s with a second order (critical) endpoint, is completely isostructural: The high-$T$ paramagnetic phases of  Cr$_x$-V$_{2-x}$O$_3$ are always associated with a corundum crystal structure.

The experimental evidence of the MIT in V$_2$O$_3$ has been accumulated, first for static quantities (e.g., the dc resistivity) and -ar a later time- for spectral functions (PES\cite{Mo2003}, ARPES\cite{Lupi2010}, XAS\cite{Park2000,Rodolakis2010}, etc...). In this paper, however, we focus on infrared-optical spectroscopy\cite{Rozenberg1995,Baldassarre2008,Lupi2010} only, which is a bulk sensitive technique in comparison to photoemission, and -contrary to XAS- includes important information about the itinerant part of the electronic properties of strongly correlated electron systems. In optical spectroscopy measurements  at room $T$, the crossing of the MIT upon Cr-doping is clearly reflected in the abrupt disappearance of the (weak) Drude peak in the in-plane\cite{nota_opt_c} optical conductivity $\sigma(\omega)$ with the opening of a sizable spectral gap. Further important information has been also extracted from the temperature\cite{Baldassarre2008} and pressure\cite{Lupi2010} dependence of $\sigma(\omega)$: The former has provided a clear indication of a strong interplay between small changes of the lattice parameters and electronic properties, while the latter (together with XAS measurements of the V K-pre-edge) has proven the inconsistency of the long-standing assumption of equivalence of doping-level and applied pressure in the phase diagram of V$_2$O$_3$.
Also to be mentioned are very recent optical measurements\cite{Lupi2010} performed in the most  ``intriguing'' region of the phase-diagram, i.e., right across the MIT first order transition line: The combined analysis of optical data and photoemission on a microscopic scale has demonstrated the formation of insulating islands embedded in the PM phase in the metallic side of the MIT. The formation of such islands, growing in size when the transition is approached, can be put -to some exert- in analogy with the nucleation processes due to impurities in a standard liquid-gas transition: In the case of V$_2$O$_3$ the impurity would be likely provided by the lattice distortions\cite{Frenkel2006} due to the Cr- substitutions.

From the theoretical point of view, the problem to be analyzed consists of a systems with two electrons in the three $3d$-$t_{2g}$ (i.e., correlated) bands of the $V$ atom at the Fermi level. The $t_{2g}$ basis further splits in one $a_{1g}$ and two $e_{g\pi}$ local orbitals (separated by $0.2$-$0.3$eV) because of a slight trigonal distortion of the material (see, e.g., LDA calculations with $N$th order muffin-tin orbitals~(NMTO) in Ref. \onlinecite{paperI}).  As clearly stated in Ref. \onlinecite{paperI}, the interplay between strong electronic correlation and multi-orbital physics is expected to be the crucial ingredient of the physics underlying the Mott MIT in in V$_2$O$_3$. In fact, the properties of the MIT in the Cr-doped V$_2$O$_3$ have been calculated (in some case even preceding the experimental results) by means of LDA+DMFT in Refs. \cite{Held2001,Keller2004}, and, later, by including the orbital hybridization in Ref. \onlinecite{Poteryaev2007}.  

Beside the success in describing photoemission data, LDA+DMFT can be also used to analyze optical spectra. While -at the DMFT level- the numerical effort for computing the optical conductivity $\sigma(\omega)$ is comparable to that for computing spectral functions, as vertex corrections can be usually neglected\cite{note_vertexcorr}, rough approximations have been always done in evaluating the optical dipole matrix elements in the localized (NMTO, Wannier, etc..) orbital basis. In particular, in the first LDA+DMFT calculations  of $\sigma(\omega)$ for V$_2$O$_3$ \cite{Baldassarre2008} the dipole matrix elements were simply replaced by $1$, while in later works\cite{Tomczak2009} the dipole matrix elements have been evaluated in the Peierls approximation,  including the effects of multiple atoms in the unit cell when necessary\cite{Tomczak2009a,Tomczak_thesis}.


 Our results for V$_2$O$_3$  are summarized in Fig. \ref{Fig:Parameter1}, where we show in the first row LDA (left) and LDA+DMFT (right) calculations for the optical conductivity obtained by using the optical matrix elements, while in the second row the corresponding calculations made with the PA are discussed~(we use $a=4.95 \AA$ , $c=14\AA$  as lattice parameters, see Ref.~\onlinecite{paperI} and references therein). In all cases, we also directly compare our theoretical results with the experimental data reported in Ref. \onlinecite{Rozenberg1996}.
Our analysis at the level of the optical conductivity clearly confirms the pivotal role played by electronic correlations in the physics of V$_2$O$_3$: the LDA results show a  much stronger Drude peak (almost an order of magnitude stronger)  when compared to the experiments. The inclusion of correlations via DMFT significantly improves the situation: Due to the proximity to the Mott-Hubbard MIT one observes a marked spectral weight shifts from the Drude peak to higher frequencies, which makes the overall agreement with experiment much better in the region up to $1.5$eV, where the experimental data are available.

From our analysis, moreover, another important aspect emerges: in the case of V$_2$O$_3$ the PA (adopted in previous calculations, e.g. \onlinecite{Lupi2010}) works satisfactorily well, at least in the low-energy $t_{2g}$ subspace: The improvements due to the inclusion of the full optical matrix elements only leads to small changes in the optical spectra up to $2$eV both in the LDA and LDA+DMFT results, as it can be expected on the basis of the discussion of Section IIB, considering the small (or vanishing) value of the crystal field splitting between the localized $t_{2g}$ orbitals at the Fermi level.


\section{Conclusion}\label{Sec:summary}

We have developed a program package for calculating the optical conductivity
on the basis of Wien2K, and make it available  to the scientific
community at {\sl www.wien2k.at/reg\_user/unsupported/wien2wannier}.
Electronic correlations, e.g., from DMFT, or finite life times, e.g., from impurity scattering, can be included via a corresponding self energy for the Wannier orbitals.
From this self-energy the Green function is calculated, which together with the full dipole matrix elements yields the optical conductivity, disregarding
vertex corrections\cite{note_vertexcorr}.

The main topic of the paper is a careful comparison between the dipole
matrix element approach and the Peierls approximation, which is the de facto standard for lattice model calculations. We have considered two materials
SrVO$_3$ and V$_2$O$_3$ as testbeds. The low frequency part (below 2-3 eV)  of the optical conductivity stems from $d$-$d$ transitions, at least for the two materials considered and many other transition metal oxides. This part is well captured by the Peierls approximation. One can understand this by  the high degree of localization of the degenerate (or almost degenerate) Wannier $d$ orbitals: Below $\sim$ 1 eV, for both vanadates, it is also sufficient to include only the three $t_{2g}$ bands out of the five $d$-orbitals.
For the high frequency part (above $2-3$ eV), on the other hand, not only the
$d$ Hubbard bands are relevant but also $p$-$d$ transitions. This part of the spectrum is not well described by the Peierls approximation. Generalized Peierls approximation, while still approximate, also
improves the description of p-d transitions~\cite{Tomczak2009a} at a computational cost 
comparable to the full dipole matrix calculation.

The comparison to experiment shows that LDA+DMFT with full dipole matrix elements gives a good description of the optical conductivity. In contrast, the Peierls approximation  shows strong deviations
at high frequencies. The same is true for the LDA optical conductivity even with the full dipole matrix elements. For instance, the LDA  optical of SrVO$_3$ conductivity particularly shows a too pronounced peak at  $\sim 3.5$eV. This peak stems from $d$-$p$ transitions, and the DMFT correctly spreads the $d$ orbital spectral over a larger energy region: Hubbard side bands are formed and the electron-electron scattering smears out the $d$ bands.

The residual differences  between LDA+DMFT and experimental infrared-spectra
hence  cannot be ascribed to the limitation of the Peierls approximation, but rather to effects beyond the LDA+DMFT scheme. For example, impurity scattering and the inclusion of non-local electronic correlations. The inclusion of the latter requires a considerable efforts of going beyond the standard LDA+DMFT scheme, e.g., by  cluster extension of DMFT\cite{Maier2007}, dynamical vertex approximation \cite{Toschi2007} or duals fermion\cite{Rubtsov2008}, which also necessarily requires a proper treatment of vertex corrections.


We thank P. Blaha,  P. Hansmann, I.A. Nekrasov, S. Biermann, J.~M. Tomczak and A. Pimenov for helpful discussions.
JK, AT and KH acknowledge financial support from the Research Unit FOR 1346
of the
Deusche Forschungsgemeinschaft and the Austrian Science Fund (FWF project
ID  I597-N16); PW was supported by the  SFB ViCoM (FWF project ID F4103-N13)
and GK W004. Calculations have been done on the Vienna Scientific Cluster~(VSC).

\onecolumngrid


\begin{thebibliography}{10}
\bibitem{note_vertexcorr} When considering single band models in DMFT, the locality of all irreducible vertices  ensures that all vertex correction in the optical conductivity exactly vanish\cite{Pruschke93a,Pruschke93b}. While this may be not rigorously true for a (realistic) multiorbital case, it remains usually a good approximations\cite{Tomczak_thesis} when the DMFT assumptions are well fulfilled.
\bibitem{footnotekmesh} The Wien2k calculations have been performed with $1000$ $\mathbf{k}$-points for SrVO$_3$ and with $4000$ $\mathbf{k}$-points for V$_2$O$_3$. For the wien2wannier projections we used $1000$ $\mathbf{k}$-points for both materials.
\end{thebibliography}
\end{document}